\documentclass{article}

\usepackage{arxiv}

\usepackage[utf8]{inputenc} % allow utf-8 input
\usepackage[T1]{fontenc}    % use 8-bit T1 fonts
\usepackage{hyperref}       % hyperlinks
\usepackage{url}            % simple URL typesetting
\usepackage{booktabs}       % professional-quality tables
\usepackage{amsfonts}       % blackboard math symbols
\usepackage{nicefrac}       % compact symbols for 1/2, etc.
\usepackage{microtype}      % microtypography
\usepackage{lipsum}
\usepackage{graphicx}
\graphicspath{ {./images/} }

\usepackage[utf8]{inputenc} % set input encoding (not needed with XeLaTeX)

\usepackage{graphicx} % support the \includegraphics command and options

\usepackage{booktabs} % for much better looking tables
\usepackage{array} % for better arrays (eg matrices) in maths
\usepackage{paralist} % very flexible & customisable lists (eg. enumerate/itemize, etc.)
\usepackage{verbatim} % adds environment for commenting out blocks of text & for better verbatim
\usepackage{subfig} % make it possible to include more than one captioned figure/table in a single float

\usepackage{enumitem} % Adds letter listing for parts ina  problem
\usepackage{physics} % adds simplified derivative and partial derivative calls (\dv[n]{}{} \pdv[n]{}{} \fdv[]{}{})
\usepackage{tikz} % Checkmarks
\usepackage{listings}
\usepackage{mathrsfs} %Fancy script letters for math symbols 
\usepackage{gensymb}
\newcommand{\beq}{\begin{equation}}
\newcommand{\eeq}{\end{equation}}
\def\bal#1\eal{\begin{align*}#1\end{align*}}

\newcommand{\bseq}{\begin{subequations}}
\newcommand{\eseq}{\end{subequations}}

\newcommand{\bi}{\begin{itemize}}
\newcommand{\ei}{\end{itemize}}
\newcommand{\I}{\item}
\newcommand{\be}{\begin{enumerate}}
\newcommand{\bea}{\begin{enumerate}[label=(\alph*)]}
\newcommand{\ee}{\end{enumerate}}
\newcommand{\bc}{\begin{center}}
\newcommand{\ec}{\end{center}}
\newcommand{\bpm}{\begin{pmatrix}}
\newcommand{\epm}{\end{pmatrix}}
\newcommand{\avg}[1]{\langle #1 \rangle}

\newcommand{\figMODEL}{
\begin{figure}[!h]
    \centering
	\includegraphics[width=0.75\linewidth]{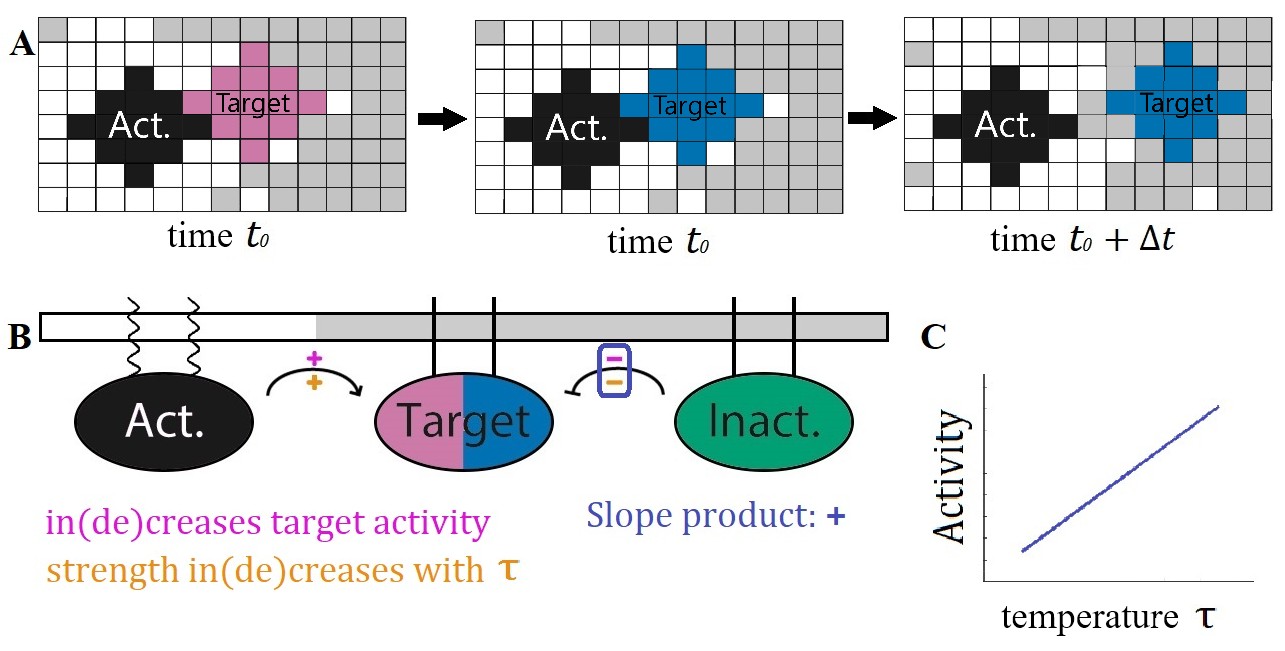}
	\caption{\textbf{Effective reaction rates between components depend on solvent's distance to the critical point.} \textbf{A:} Signal components are disks of spins that diffuse in the 2D Ising Model. At time $t_0$, an activator (black) comes in contact with an inactive target (pink). In the same time step, the target is activated (blue) at some fixed rate. During the next sweep, the activated target diffuses away. Because of coupling to Ising degrees of freedom, the effective rate of these reactions depends on the state of the surrounding membrane through their rate of coming into contact. \textbf{B:} Cascade diagram with component partitioning. Sign above reaction arrow (purple +/-) indicates that the pathway increases/decreases target activity. Sign below reaction arrow (orange +/-) indicates whether interaction rate increases or decreases with rescaled temperature.  The product of these two signs (here positive in both cases) predicts how increasing temperature changes the target's activity. \textbf{C:} In this PPI the product of the upper and lower signs is positive for all reactions, indicating target activity should increase monotonically with respect to temperature, as sketched.  While the explicit reaction rates do not change with temperature, the effective rate of the activating reaction increases while the deactivating reaction decreases as unlike components interact more at higher temperature and like components interact less.}
	\label{fig:model}
\end{figure}
}

\newcommand{\figSTATIC}{
\begin{figure}[!h]
    \centering
	\includegraphics[width=0.75\linewidth]{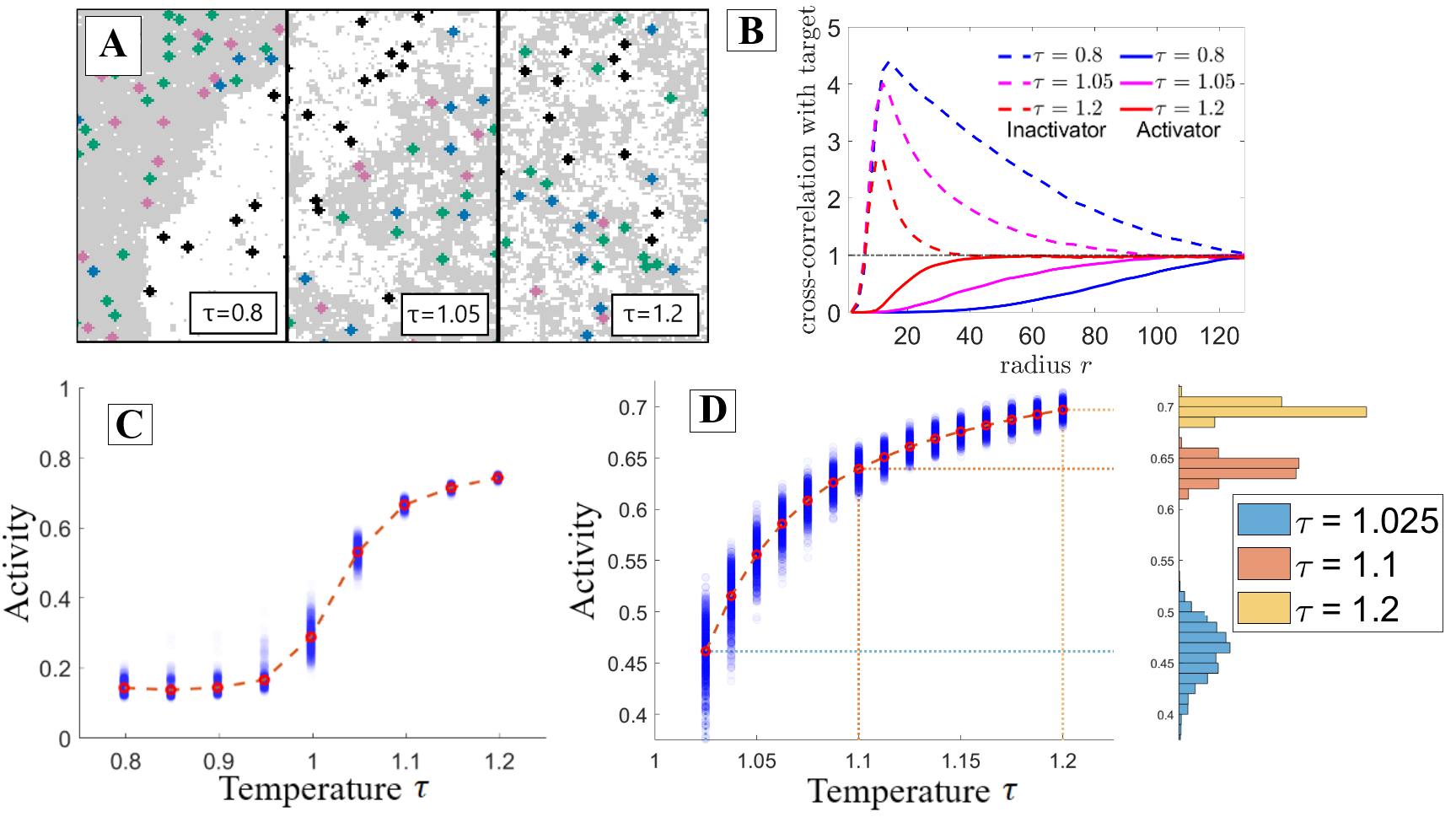}
	\caption{\textbf{Interaction networks are highly sensitive to solvent properties near criticality.}  Results are from the interaction network depicted in figure \ref{fig:model}B when carried out with Monte Carlo simulation. \textbf{A:} Configurations of the simulation at a range of rescaled temperatures. As temperature increases domains become smaller and the target is more likely to be found near its activator which prefers a different domain, thus increasing the activity. \textbf{B:} Radially-averaged cross-correlation functions between the activators and targets (solid lines) and between the inactivators and targets (dashed lines) for the three temperature values in part \textbf{A}. As temperature is increased, the activator is found closer to the target while the inactivator is found more distant. \textbf{C:} The target's activity grows with rescaled temperature, most steeply near the critical point. \textbf{D:} Taking a closer look at the near-critical region of part \textbf{C}, the greatest increase in target activity occurs between $\tau=1$ and $\tau=1.1$.  Binning the various simulation runs by activity clearly demonstrates the increased variance near the critical point. }
	\label{fig:eq_cascade}
\end{figure}
}

\newcommand{\figDISKSIZE}{
\begin{figure}[!h]
    \centering
    \includegraphics[width=0.75\linewidth]{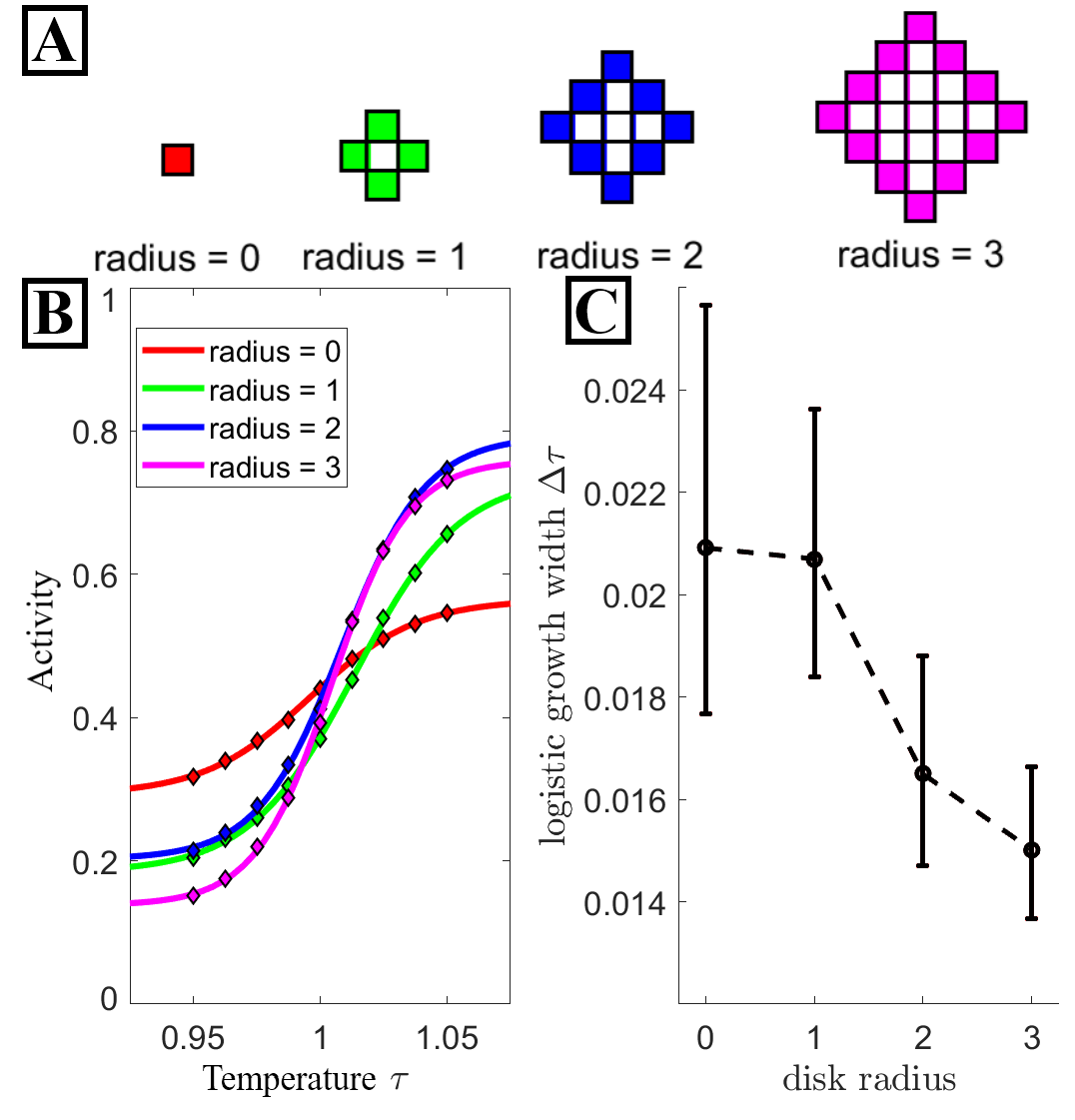}
    \caption{\textbf{Larger components are more sensitive to critical solvent properties.}   %Impact of varying the disk size for the network depicted in figure \ref{fig:model}B. 
    \textbf{A:} Visual schematic of lattice shape for each disk radii discussed here. 
    A network containing radii zero components has spatial configurations equivalent to the standard 2D Ising model. The solid boundary of each disk denotes the spins which interact with the solvent while the spins they bound have no impact. \textbf{B:} Activity data (diamond markers) for various disk radii along with the result of fitting each data set with a logistic function (solid lines).  \textbf{C:} The growth width $\Delta \tau$ of the logistic curve fits from \textbf{A}. Increasing disk size results in a stronger sensor of $T_c$. The error bars here denote the 95\% confidence interval for each $\Delta\tau$ value. All fit values are tabled in the supplement.}
    \label{fig:disk_size}
\end{figure}
}

\newcommand{\figNONZEROMAG}{
\begin{figure}[!h]
    \centering
	\includegraphics[width=0.75\linewidth]{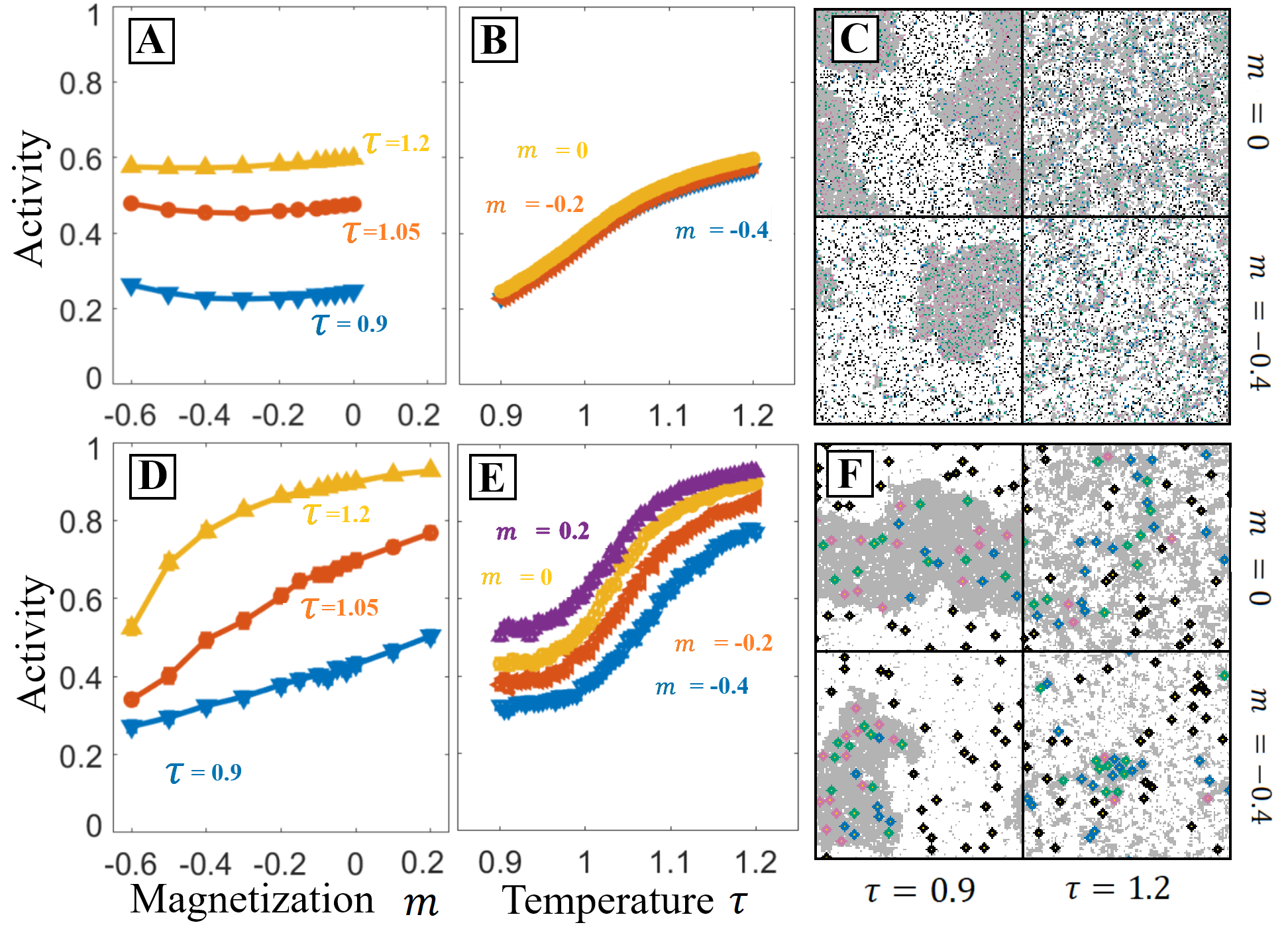}
	\caption{\textbf{PPIs with large components are also sensitive to the composition parameter parallel to the tie-lines.}  In the Ising model, this is parameterized by magnetization, $m$, which previous figures take to be zero (equal area of bright and dark phases below $T_c$).
       Here we explore varying $m$ for systems with small components ($r=0$, \textbf{A}-\textbf{C}) and larger components ($r=3$ \textbf{D}-\textbf{F}), for the interaction network depicted in figure \ref{fig:model}B. \textbf{A,B}: Activity as a function of $m$ for three values of temperature and activity as a function of temperature for three values of $m$. When components are small results do not depend strongly on $m$, though they do depend on temperature. %\textbf{B}:Activity as a function of temperature. When interaction network contains radii $0$ components, variation in $m$ has little-to-no impact on the activity.
       \textbf{C}: configurations at two values of $m$ and $\tau$.  At negative $m$, $\tau<1$ the system has less dark phase.  Above the critical temperature, there are still fewer dark domains.  \textbf{D,E}: Activity as a function of $m$ and $\tau$ for the same network with larger components, $r=3$. With these larger components the system is sensitive to both $m$ and $\tau$ near the critical temperature.  \textbf{F}: At negative $m$, there is less dark phase similar to \textbf{C}, but the larger inclusions form higher contrast domains even above the critical temperature of the bare Ising model.}
	\label{fig:solvent_comp}
\end{figure}
}

\newcommand{\figDYNAMIC}{
\begin{figure}[!h]
    \centering
	\includegraphics[width=0.75\linewidth]{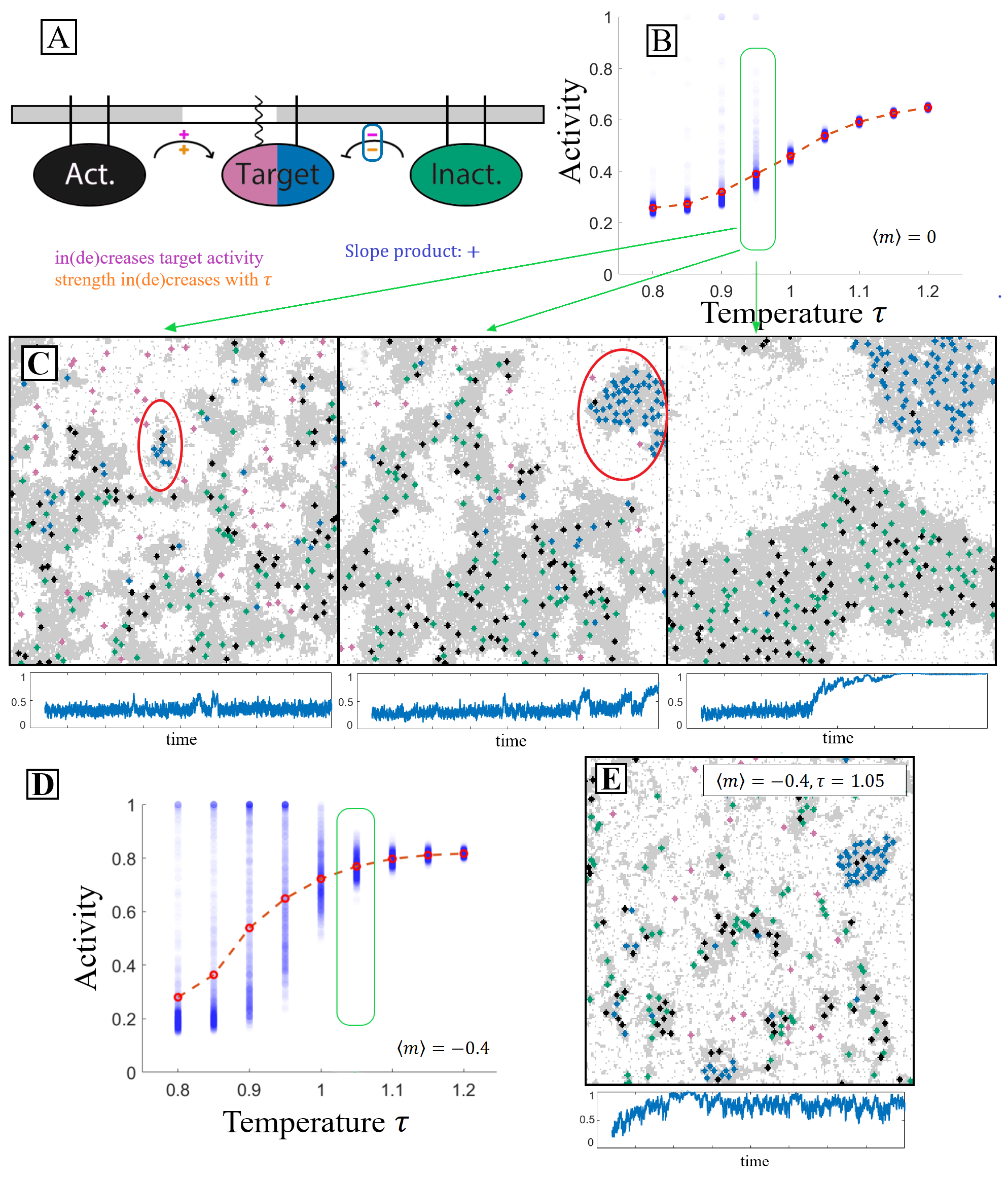}
	\caption{\textbf{Dynamics that modify component partitioning produce non-equilibrium domains.} \textbf{A}: Diagram of an interaction network in which the target's partitioning is modified alongside its state. The target’s inactive state (pink) prefers disorder while its active state (blue) prefers order indicated by the respective membrane legs. Qualitative analysis suggest the target’s activity should increase with $\tau$. \textbf{B}: Mean activity measurements follow these predictions, but below the critical temperature a small number of simulations produce outliers with very high activity. We term this phenomenon pocketing and trace it to the formation of non-equilibrium domains. \textbf{C}: Three example configurations from simulations in which pockets form with their activity vs time traces below.  We understand pocket formation kinetically. First, a pocket originates when an activator (dark preferring) and an inactive target (white preferring) meet in a disordered (white) domain, leading to target activation.  Now each prefer the dark (ordered) domain and with some probability such a dark domain will form around them before they diffuse to a larger dark domain.  We term this domain a pocket.  Once this pocket is formed, inactive targets (white preferring) can diffuse through the surrounding white phase. But when they make contact with the pocket, they can be activated, preventing them from leaving. Dark lipids are small enough to diffuse through the white phase to join the pocket as well but inactivators are excluded kinetically.  \textbf{D,E}: The same interaction network in a system with composition adjusted to $m=-0.4$ demonstrates an enhanced pocketing rate, including at temperatures $\tau > 1$. }
	\label{fig:non_eq_cascade}
\end{figure}
}

\newcommand{\figSUPPDISKSIZE}{
\begin{figure}[!h]
    \centering
    \includegraphics[width=0.75\linewidth]{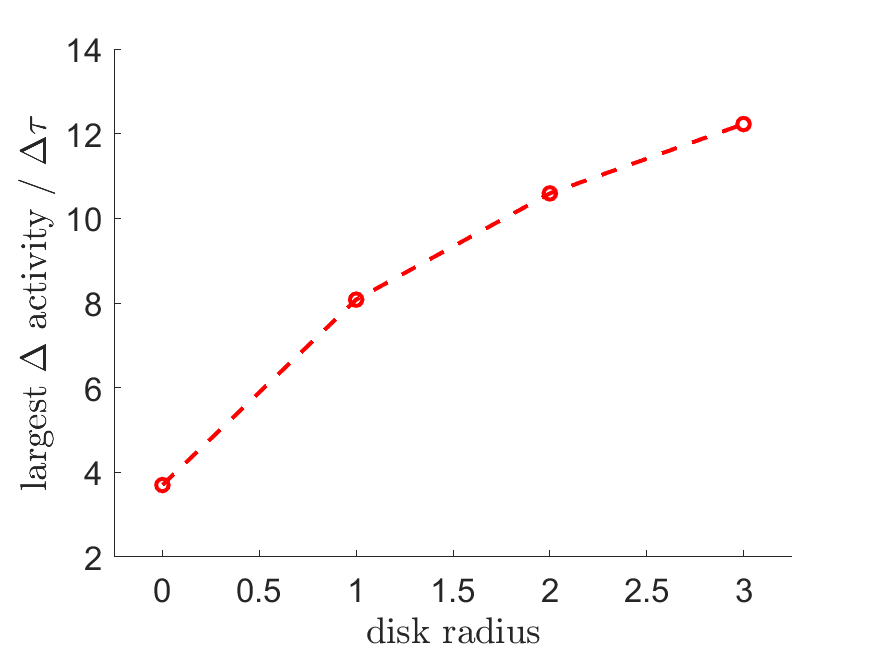}
    \label{fig:SUPP_disk_size}
    \caption{An alternative view the of activity's sensitivity gain as a function of component radius can be observed by looking at the maximum slope of the activity vs temperature for each radii.}
\end{figure}
}

\usepackage{lipsum}

\title{Protein-protein interaction networks can be highly sensitive to the membrane phase transition}

\author{
 Taylor Schaffner \\
  Department of Physics\\
  Yale University\\
  New Haven, CT \\
  \texttt{taylor.schaffner@yale.edu} \\
    \And
 Benjamin Machta \\
  Department of Physics\\
  Yale University\\
  New Haven, CT \\
  \texttt{benjamin.machta@yale.edu} \\
}

\begin{document}
\maketitle

% \begin{frontmatter}
\begin{abstract}
Many protein-protein interaction (PPI) networks take place in the fluid yet structured plasma membrane. Lipid domains, sometimes termed rafts, have been implicated in the functioning of various membrane-bound signaling processes. Here, we present a model and a Monte Carlo simulation framework to investigate how changes in the domain size that arise from perturbations to membrane criticality can lead to changes in the rate of interactions among components, leading to altered outcomes.  For simple PPI networks, we show that the activity can be highly sensitive to thermodynamic parameters near the critical point of the membrane phase transition.  When protein-protein interactions change the partitioning of some components, our system sometimes forms out of equilibrium domains we term pockets, driven by a mixture of thermodynamic interactions and kinetic sorting. More generally, we predict that near the critical point many different PPI networks will have their outcomes depend sensitively on perturbations that influence critical behavior.
\end{abstract}
% \input{content/significance}
% \end{frontmatter}

\section*{Introduction}

Once thought of as a uniform two-dimensional solvent for membrane-bound proteins, the plasma membrane is recognized today as having a heterogeneous yet fluid structure. The spatial organization in the membrane is likely due in part to membrane component interactions with the actin cytoskeletal network \cite{SHI20181769, PhysRevX.7.011031, Wulfing2266}, inter-protein interactions \cite{DOUGLASS2005937, Janosi2012}, and membrane morphology \cite{Jarsch2016, B608631D,Cui2011}. But in addition, spatial organization arises directly from thermodynamic interactions between lipids through proximity to a miscibility critical point.  Liquid membranes have a rich phase behavior. Synthetic giant unilamellar vesicles (GUVs) exhibit miscibility critical points with critical fluctuations consistent with the 2D Ising model \cite{Veatch17650, HONERKAMPSMITH2008236}. Vesicles extracted via blebbing from live cells demonstrate critical behavior without experimental fine-tuning of composition, suggesting that \emph{in vivo} their membranes may be poised close to criticality \cite{Baumgart3165, doi:10.1021/cb800012x}.  Extrapolation to live cells seems to indicate that intact membranes are tuned slightly above the critical point where there is no macroscopic phase separation, but where relatively large domains up to a correlation length of$~20 \textrm{ nm}$ are expected. Although the thermodynamics of the membrane are now better understood, we do not yet understand how proximity to criticality impacts the protein interaction networks that take place in the two-dimensional solvent of the plasma membrane.

We have previously explored the ramifications of this proximity to criticality. We and others have shown that pairs of proteins embedded in a nearly critical solvent may feel effective forces mediated by thermodynamic domains that can extend to a correlation length, much larger than electrostatic interactions \cite{Machta_2012, Shelby2023-yo, Katira2016, Reynwar2008}. We also showed that single ion channels that interact with lipids can be sensitively regulated by changes to their solvent environment near demixing criticality \cite{Kimchi2018, Suma2024}. An area that remains unexplored is how the membrane coupling could impact the protein-protein interactions that take place in the membrane. Here, we seek to understand how membrane domains can alter which components interact with each other, thereby shaping the outcomes of protein interaction networks and signaling cascades.  

Diverse lines of evidence suggest that protein-protein interaction networks are sensitive to their membrane environment.  Perturbations known to influence the distance to criticality often impact PPI networks.  Miscibility critical points have two sensitive directions; here, area fraction of liquid-ordered domains and the distance to the critical temperature which sets the typical size of domains. Cholesterol depletion and loading primarily change the area fraction of domains and impact many signaling assays, including that in neurotransmitters \cite{Allen2007} and RAS clusters \cite{Janosi2012}. We have shown that conditions that induce and stabilize domain formation by lowering transition temperature also reverse the anesthetic effects of short chain n-alcohols \cite{Machta2016}.  In addition, many components of protein-protein interaction networks have been shown to partition into membrane domains. In B lymphocytes, activating kinases often partition into ordered domains, while deactivating phosphatases partition into disordered domains \cite{Stone2017}. Furthermore, some PPI networks lead to downstream changes in domain size and composition \cite{Linder2007}; for example, B-Cell receptor (BCR) activation increases ordered domain formation \cite{Stone2017}.

In this manuscript, we demonstrate that when signaling components couple to different domains near a critical point, signaling outcomes can sensitively depend on thermodynamic parameters that affect the stability of domains. We present a Monte Carlo simulation framework to probe the robustness of this behavior alongside a diagrammatic method of predicting changes in signaling outcomes. We find that interactions between larger proteins become especially sensitive to perturbations that change the critical temperature and thus the domain size. We additionally explore mechanisms that modify domain preference, like palmitoylation, and find that interaction networks that include such reactions can produce far from equilibrium domains.  

\section*{Methods}

\subsection*{Model}
\figMODEL
Here we develop a minimal simulation framework to investigate PPIs solvated by a membrane poised close to a liquid-liquid critical point. We represent the membrane with a 2D square lattice. We represent membrane lipids as Ising spins $s_i = \pm 1$ \cite{Machta2011}, corresponding to components that preferentially partition into the ordered/disordered phases of real membranes. Two lines of experimental evidence suggest equating our lattice spacing with approximately $1$nm; extrapolating the correlation length near the critical point of cell-derived vesicles yields an $l$ near $2\textrm{ nm}$ \cite{doi:10.1021/cb800012x}, and the size of a membrane lipid generally falls around $~0.8 \textrm{ nm}^2$ \cite{doi:10.1016/S0006-3495(02)73949-0}. 

Membrane-embedded components of PPIs in our model, such as kinases and phosphatases, are represented as disk-shaped inclusions. As in previous work \cite{Kimchi2018}, these inclusions are coupled to the membrane in which they are embedded by coupling to neighboring lipids' Ising spins.  These disk-shaped inclusions are bounded by uniform sets of spins $\{b\} _{i}$ where $i$ denotes the $i^{\textrm{th}}$ inclusion in the membrane. The Hamiltonian of this system has the form
\beq
    H_{\textrm{tot}} = H_{\textrm{mem}}(\{s\}) + \sum^{N}_{i=1} H_{\textrm{int}}(\{s\}_{\partial}, \{b\}_i)
    \label{eq:hamiltonian}
\eeq
where $H_{\textrm{mem}}$ contains the energy of membranes lipids' interacting among themselves and $H_{\textrm{int}}$ represents the interaction of the spins on the boundaries of inclusions with their adjacent membrane spins denoted by $\{s\}_{\partial}$.

To capture the diffusive dynamics of membrane lipids, we use the Kawasaki algorithm in which spins are swapped with their neighbors with a probability chosen to obey detailed balance.  These dynamics locally conserve the order parameter and are thus in the model B universality class \cite{RevModPhys.49.435}. Our model does not include hydrodynamic interactions that may contribute at cellular length and time scales \cite{HONERKAMPSMITH2008236}. Inclusions translate to neighboring lattice sites with moves that obey detailed balance \cite{Tasios2016}, simulating the diffusive dynamics of proteins. In a standard sweep of the algorithm, an attempt is made to move every lipid on the lattice, and $C$ attempts are made to move each disk (see supplement for details).

The interactions between the PPI components are carried out as shown in figure \ref{fig:model}A. When two particular components come into contact, a reaction can change the state of one, reflecting a biochemical modification such as phosphorylation. Because the interacting partners must be in contact, the effective reaction rates thus depend implicitly on the properties of the nearly critical solvent in which they are embedded.  These dynamics represent non-equilibrium reactions, and thus do not satisfy detailed balance.

\subsection*{Qualitative Analysis}

By detailing the interaction network's components' partitioning along with their interactions as shown in figure \ref{fig:model}B, a straightforward prediction may be made about the PPI's outcome as the system passes through criticality. This particular interaction network contains the chemical reactions 
\bal
    A + T_{I} &\rightarrow A + T_{A} \\
    I + T_{A} &\rightarrow I + T_{I}
\eal
where $A$ denotes the activator (black), $I$ denotes the inactivator (green), and $T_{I/A}$ denotes the inactive/active target (pink/blue). Each component's partitioning is denoted by the legs that attach it to the membrane from above--wavy lines for $l_\text{d}$ and straight for $l_\text{o}$.  On each pathway, denoted as arrows in the diagram, two signs are labeled. The sign above the pathway refers to if this interaction increases or decreases the activity of the component of interest, denoted here as the target. The orange sign below the pathway indicates if the rate of contact between the two components increases or decreases with the rescaled temperature $\tau$ where

$$ \tau = \frac{T}{T_\textrm{c}}.$$ 

In cells, this rescaled temperature could be accessed either by changing temperature $T$ or through changes in composition which change the critical temperature $T_c$.
This orange sign is negative if the two components partition into the same domains and positive if they partition into different domains. If the product of the upper and lower signs on each pathway agrees for all pathways in the interaction network then we expect this product to give the sign of the slope of the target's activity with respect to $\tau$ as shown in figure \ref{fig:model}C. This scheme allows for ease of creating interaction networks with straightforward predictions from our model.

\section*{Results}

\subsection*{Domains regulate protein-protein interactions according to partitioning}
\figSTATIC
We first wanted to understand how solvent properties near the critical point could influence PPIs whose components have fixed partitioning into domains. In these systems, the configurations of proteins and lipids are in equilibrium, so that the probability of a configurational state is given by a Boltzmann weight. However, while spatial configurations are given by this equilibrium distribution, the dynamics of the activity state of target proteins is not in equilibrium.

As a model of such a system, we explored the cascade depicted in Figure \ref{fig:model}\textrm{B}. Here the receptor state that we are interested in tracking is denoted as the target which has an inactive (pink) state and an active (blue) state. Alongside the target in this cascade, there are inactivating (green) and activating (black) components. The target and the inactivating component partition into the ordered domain while the activating component partitions into the disordered domain. In each simulation, the components and the lipids are allowed to evolve under locally conserved Kawasaki dynamics, and the target activation rate achieves steady-state.

The model predictions may be understood visually by analyzing how the spatial organization of components modifies their respective interaction rates. In Figure \ref{fig:eq_cascade}\textbf{A}, three snapshots show simulations at different $\tau$. As $\tau$ increases, the oppositely-partitioned activators and targets come into contact more frequently, resulting in increased activity. The spatial correlation analysis shown in Figure \ref{fig:eq_cascade}B demonstrates the correlation change between components. As temperature increases, the target is more often localized near the activator compared to the inactivator. 

Figure \ref{fig:eq_cascade}\textbf{C} demonstrates that as predicted in figure \ref{fig:model}\textbf{C}, the target's activity in this protein-protein interaction network increases as the rescaled temperature increases. Additionally, we observe the greatest changes in activity, and thus effective interaction rate among PPI components, near the thermodynamic critical point. This is because near the critical point the typical size of domains becomes sharply dependent on the thermodynamic rescaled temperature, either through actual temperature, or through the critical temperature. A finer look at activity as a function of temperature is shown in figure \ref{fig:eq_cascade}\textbf{D}. Near the critical point, the activity becomes highly susceptible to perturbations in the rescaled temperature.

%\subsection*{Sensitivity increases with component size}
\subsection*{Sensitivity increases with PPI component size}
\figDISKSIZE
Inspection of the Hamiltonian in Eq. \ref{eq:hamiltonian} reveals that the energy contribution per inclusion increases with inclusion size.  In Fig. \ref{fig:disk_size}, we explore how target activity changes as we vary PPI component radius for the same interaction network. In Fig. \ref{fig:disk_size}\textbf{A} we show the lattice shapes of the disks for which we plot data. PPI's containing larger components are more sensitive to temperature changes (see Fig. \ref{fig:disk_size}\textbf{B}). To quantify these curves, we fit the fraction of active targets $f_A$ in the steady-state system to a generalized logistic function of temperature,

\beq
    f_A (\tau) = \frac{R - L}{2} \tanh{ \frac{(\tau - \tau_0^{\textrm{fit}})}{2\Delta \tau} } + \frac{R+L}{2}, 
    \label{eq:sigmoid_eq}
\eeq
where the fit parameters $L$ and $R$ denote left and right asymptotes, respectively, $\Delta\tau$ the growth width, and $\tau_0^{\textrm{fit}}$ is the point of steepest activity increase.  We note that $\tau_0^{\textrm{fit}}$ is near 1, consistent with the idea that near the critical point is where PPIs are most sensitive to solvent properties.  We also note that larger proteins have the effect of raising the critical temperature slightly, and we report $\tau$ for a bare membrane, so that the true thermodynamic critical point in these systems is at $\tau$ slightly larger than 1, increasing for larger inclusions and larger inclusion density.

The fitted values for the growth width $\Delta \tau$ are plotted in Fig. \ref{fig:disk_size}\textbf{C} for each radius. As intuited, the growth width decreases with increasing radius, demonstrating that sensitivity to $\tau$ increases with the size of the PPI component.

%\subsection*{Large proteins can be sensitive to the area fraction of different domains.}
\subsection*{Large proteins are sensitive to domain area fraction}
\figNONZEROMAG
The results we have presented thus far have considered how network activity is modulated by changes to rescaled temperature. This changes the size and contrast of domains, but without altering the relative ratio of bright to dark domain area, moving perpendicular to the low temperature tie lines. Here we explore how changes to the ratio of these two domain areas impact PPIs, moving parallel to these low temperature tie lines.  In the Ising model we can access this by changing the relative fraction numbers of spins with the value $s_i=1$, quantified through the magnetization $m=<s_i>$.  $m \neq 0$ corresponds to membranes with unequal area fractions of liquid-ordered and liquid-disordered phases when cooled below the phase transition temperature.  This effective parameter depends on membrane composition, and can be accessed by depleting and loading cholesterol, or more broadly by any perturbation which moves along a tie line in the phase separated regime. In Fig. \ref{fig:solvent_comp}, we look at how varying $m$ and $\tau$ impact the same PPI detailed in Fig. \ref{fig:model} for a range of component sizes.

Fig. \ref{fig:solvent_comp}\textbf{A} and Fig. \ref{fig:solvent_comp}\textbf{B} demonstrate that variation in $m$ has very little impact on the activity for the PPI with components of radii $0$. Conversely, Fig. \ref{fig:solvent_comp}\textbf{D} and Fig. \ref{fig:solvent_comp}\textbf{E} indicate that varying $m$ has a substantial effect on the activity of a PPI with components of radii $r=3$. As $m$ decreases, the frequency of interaction between the like-partitioned inactivator and target increases while the interaction between the opposite-partitioned activator and target decreases. The resulting effect is a monotonic increase in the target's activity with $m$ that continues across the range of negative to positive $m$. Comparing the 4-panels in Fig. \ref{fig:solvent_comp}\textbf{C} to Fig. \ref{fig:solvent_comp}\textbf{F} suggests that this difference is in part due to changes in the system's spatial organization that arise from the varying strength of the force that causes solvent to aggregate around extended objects \cite{Tasios2016, Katira2016}. In particular, the $m=-0.4$, $\tau=1.2$ panel demonstrates high contrast domains which are absent in simulations with radius 0 proteins.

\subsection*{Coupling protein partitioning to PPI dynamics can drive non-equilibrium domain formation}

\figDYNAMIC

The results we have presented so far have domains and spatial fluctuations that are in thermal equilibrium.  While the activity state of the proteins is out-of-equilibrium and does not satisfy detailed balance, but because these internal states do not feed back onto configurational degrees of freedom, spatial configurations are exactly given by the Boltzmann distribution.  Here we consider PPIs wherein the activity state of proteins alters their energetic interactions with boundary lipids such that the configurations are no longer in thermal equilibrium.

We are inspired by reactions such as palmitoylation, in which fatty acids are covalently attached to proteins by specific enzymes in a reversible and signal-dependent manner \cite{ROCKS2010458, Linder2007}. Palmitoylation is a major driver of partitioning into liquid-ordered phases \cite{Lorent2017-bd, Fukata2010, Jiang2018}, indicating that these reactions likely change the domain partitioning of target proteins.  We implement similar dynamics by having target proteins change their energetic interactions with neighboring lipids in response to changes to their PPI state.  Because the rate of such reactions does not explicitly depend on the spatial configurational state of bounding lipids, this reaction scheme violates detailed balance.

A PPI that contains a domain-changing mechanism is detailed in the diagram in Figure \ref{fig:non_eq_cascade}\textbf{A}. The activator and in-activator both partition into the ordered domain. The target, when inactive, partitions into the disordered domain and upon activation changes to the ordered domain. Our qualitative analysis, which successfully predicted previous qualitative results makes a clear prediction for how the temperature should impact activity.  Because the signs on both pathways present in this interaction network have a positive product, we anticipate that the target's activity will increase with temperature. 

Aligning with the diagrammatic predictions, Figure \ref{fig:non_eq_cascade}\textbf{B} demonstrates that activity on average increases with temperature in this PPI. In addition to the average along the red-dotted line, this plot also includes the individual average activity of each of the many separate simulations of this network performed in this analysis as blue semi-transparent markers. When looking at the low-temperature clusters of these simulations, a peculiarity arises that is not present in the static partitioning case. Small subsets of low-temperature simulations exhibit extremely high levels of activity that would appear to break the predictions of our diagrams. The second and third panels of Fig. \ref{fig:non_eq_cascade}\textbf{C} contain prominent examples of what occurs in these anamolous simulations. We label this emergent behavior `pocketing'.

Simulation runs that contain pocketing have two distinct dark domain types--something not observed in equilibrium simulations. The first type corresponds to that predicted by the diagrams and equilibrium; it contains activators and inactivators, but few targets. A second dark domain type--the pocket (circled in red in Figure \ref{fig:non_eq_cascade}\textbf{C})--is filled with activated targets and contains several activators, but no inactivators. Because activators modify targets to prefer dark domains, here we term these `recruiters', and we term inactivators `excluders'.

Pocketing is an explicitly kinetic process that is best understood through its formation process. Pocketing begins when a recruiter sits in a small dark domain which contains no excluders. Inactive targets can diffuse into this domain where they become activated, trapping them in the domain. In this way, the domain grows.  Inactive targets can diffuse through the bright phase, but once activated (recruited) they cannot leave. Excluders are unable to enter the pocket because doing so would require them to cross through a bright domain. As the pocket grows, so does the energy barrier that blocks excluders and recruiters from entering and leaving, increasing its stability. We note that pocket formation frequency is dependent on details that are slow to reach steady state, an issue we discuss in more detail in the supplement

In Fig. \ref{fig:non_eq_cascade}\textbf{D} we demonstrate that the frequency with which this bistable aggregation-like behavior occurs may be considerably increased by using the aggregation-enhancing effect of increasing the ratio of inclusions to lipids in a domain as demonstrated in the domain area fraction perturbation results. In Fig. \ref{fig:non_eq_cascade}\textbf{E}, the lipid budget is tighter, i.e. when there are less ordered lipids to accumulate around order-preferring inclusions, partition-modifying dynamics have greater impact, forming and dissolving energy barriers with higher frequency and greater amplitude.

\section*{Discussion}

Here we present a model and simulation framework for protein interaction networks in a membrane near a de-mixing transition. In our model, proteins have preference for local lipid environments through their boundary. Near the critical point, small changes in composition or temperature alter the frequency with which components co-locate, changing effective reaction rates and network outcomes. We provide a diagrammatic framework to qualitatively predict how changing distance to the phase transition should impact protein interactions.  In simulation, we show that near the critical point the activity of specific targets can vary dramatically as the distance to the critical temperature is varied.  This effect is especially large for large proteins, which also become sensitive to composition parameters which do not change the critical temperature.  Finally, when some interactions modify a component's partitioning, we find that some of our simulations produce distinct out-of-equilibrium domains which we term pockets.

Our model is motivated by the findings that membranes from a range of animal cell types have compositions tuned near but above a liquid-liquid critical point \cite{Baumgart3165,Veatch17650,LI2020758}.  Different membrane bound proteins have differing partitioning into low temperature domains, presumably changing their rate of interactions with other components, even above $T_c$. On the functional side, perturbations that influence the membrane phase transition, such as n-alcohols \cite{Heimberg2007,Gray2013} and cholesterol loading/depletion \cite{Keller2000,Levental2009} often lead to changes in signaling activity.  Past work has argued that proximity to the membrane phase transition could influence membrane-mediated forces \cite{Machta_2012}, or the allosteric states of proteins \cite{Kimchi2018}, but here we show that these changes could also influence protein-protein interactions by changing the frequency with which components come into contact.

% Point: Near the critical point, PPIs are most sensitive to changes in the distance to the critical point.
The protein interaction networks that we examine here are especially sensitive to small changes in membrane properties when those membranes are tuned near their critical point. In our simulations, the reaction rates between different components are implicitly set by the rate at which they come into contact. This requires proteins to co-localize to the same local lipid environment, and the availability of different environments becomes strongly dependent on the two Ising parameters of reduced temperature ($T-T_c$) and magnetization (the fraction of dark spins). For example, reactions between pairs of proteins that prefer different low-temperature phases become strongly suppressed on lowering temperature, as their preferred environments become more different. Interaction networks that carry out signal transduction seem to use this motif. In B-cell signaling, liquid-ordered membrane domains cluster around ligand-bound receptors, enriched in activating kinase Lyn and depleted of inhibitory phosphatase CD45 \cite{Stone2017,Shelby2023-yo}.  Signaling is inhibited by n-alcohol treatments which lower critical temperatures (raise rescaled temperature), possibly by altering the balance between activating and inhibiting reactions through this mechanism \cite{Machta2016}.

%Point: cells could regulate protein interaction networks by changing membrane composition 
Our results suggest that cells could exert control over many protein interaction networks at once by altering the composition of the membrane near the critical point.  In support of this idea, experimental work in macrophage cells demonstrate that pro-inflammatory perturbations including interferon $\gamma$, Kdo 2-Lipid A, and lipopolysaccharide move the cell's membrane closer to the critical point, whereas an anti-inflammatory perturbation in the form of interleukin 4 moves the membrane away from the critical point \cite{Cammarota2020}.  Cells also modulate the transition temperature of their membrane based on growth conditions and cell density in culture, and with phase in the cell cycle~\cite{Gray2015}. Recent work also showed that cholesterol chemical potential was higher in metastatic cells than in closely related non-metastatic cell lines, and found that artificially raising cholesterol chemical potential leads to increased STAT3 phosphorylation, a pro-oncogenic signal \cite{Ayuyan2018}. These changes to the membrane solvent environment could exert functional influence through other mechanisms including allosterically modulating functional states of individual proteins~\cite{Kimchi2018,Levental2023,Mcgraw2019}, and changing the stability of prewet protein liquid domains~\cite{Rouches2021}. However, our results suggest that changes in lipid composition could also exert control more directly, by changing the outcomes of signal transduction cascades and other protein-protein interaction networks.

% \subsection*{Dynamic component partitioning as a cluster promotion mechanism}
% Point: PPIs that result in a partition change are capable of nucleating distinct phases
In simulations where component partitioning changes with protein state, we sometimes see far-from-equilibrium domains form, generated through a combination of thermodynamic interactions and kinetically driven sorting. These domains, which we term pockets, can be seen by the differential partitioning of active targets into pockets (blue components in figure 5C) even while inactivators (green) are excluded, even though they have the same interactions with surrounding lipids. In our simulations, pocketing requires that activators are recruiters, modifying their target in a way that gives them a thermodynamic preference for the local domain.  Palmitoylation \cite{Fukata2010}, in which a fatty acid is covalently attached to an amino acid, could play this role. This modification is reversible in cells, and has been shown to shift partitioning into liquid-ordered domains \cite{Levental2009,Lorent2017-bd}. A similar motif is also seen in some signaling cascades. For example, during TCR signaling, TCR and bound ZAP70 phosphorylate LAT, leading to the formation of a LAT cluster \cite{Su2016,James2012}. Here, TCR/ZAP70 act as a recruiter for the target LAT.  Different from our simulation, the LAT phosphorylation likely changes the interaction with Grb2 and SOS1, rather than through the membrane, suggesting these domains are prewet \cite{Rouches2021, Bagheri2024}.  Nevertheless, the motif of recruiting targets to a domain formed through a combination of thermodynamic forces and non-equilibrium sorting seems shared.    

%relevance in 3D
Although our focus is on protein interaction networks that take place on the two-dimensional plasma membrane, it is likely that thermodynamic domains stabilized by thermodynamic interactions between proteins lead to similar effects in three dimensions. Currently, there are many structures attributed to liquid-liquid phase separation of disordered domains within the nucleus and cytoplasm, many of which wet and prewet onto membranes \cite{Rouches2021,James2012}, chromosomes \cite{Rouches2025}, and likely other surfaces \cite{Riback2020}. In some condensate-driven systems such as TCR signaling via pLAT, Grb2, and Sos1, a combination of membrane-bound and free components work together directly linking the two- and three-dimensional systems \cite{Ditlev2021}. Away from surfaces, phase separation in many biological condensates has been ascribed to serve a wide mileu of functions including the enhancement or suppression of biochemical reactions \cite{Lyon2021}, and the means by which this may function \emph{in vivo}, as in our results, is by regulating interaction network component proximity--a feature that readily extends from two dimensions to three.

% \subsection*{PPI networks considered here could be used to tune cell membranes to their critical point}
Self-organized criticality has been proposed as a mechanism for maintaining a system near a precise operating point without fine-tuning \cite{BakSOC1988, Camalet2000, Levina2007, Magnasco2009, Landmann2021, Graf2024, Choi2024, Momi2025, Sherry2025}. Cell membranes seem to be precisely tuned near to a miscibility critical point, but the mechanism which achieves this is unknown. An investigation of zebrafish cell membranes demonstrated that they adapt their miscibility temperature to stay a fixed distance above their growth temperature \cite{Burns2017}. Distantly related yeast cells also regulate the miscibility temperature of vacuole organelle membrane to maintain a fixed distance below their growth temperature \cite{Leveille2022}. In each case, cells are thought to adjust their lipid composition to achieve this tuning. While mechanisms for lipid homeostasis are diverse, the robustness of criticality to temperature changes in these systems suggests involvement of a sensor that can detect and feedback on the distance to the critical point. Our results suggest that protein-protein interaction networks may provide such a sensor. The activity of networks that are embedded in the membrane provide a natural way to read out the solvent properties of the membrane. One aim of future work with this model will explore the development of such self-organizing systems.

\section*{Data availability}
The simulations here were carried out using a monte carlo framework developed and available at \url{https://gitlab.com/taylor-schaffner/memtropic}.

Post-simulation data analysis was carried out using MATLAB (The MathWorks). Analysis code and data files can be found organized by figure at \url{https://gitlab.com/taylor-schaffner/ppi-paper-figure-data}.

\section*{Author Contributions}

Machta conceptualized the research. Schaffner developed the computational framework, carried out all simulations, and analyzed the data. Schaffner and Machta interpreted simulation results connecting them to biological observations. Schaffner and Machta wrote the article. 

\section*{Acknowledgments}
This work was supported by NIH R35 GM138341.  We thank Yu Fu, Isabella Graf, Mason Rouches, Sarah Veatch for useful discussions. 

\bibliographystyle{unsrt} 
\bibliography{main.bib}

%%%%% Add Supplement to the end %%%%%
\clearpage % Starts a new page
\appendix % Starts the appendix section
% supplement.tex
\onecolumn

\section*{Supplementary Information}

\section{Simulation Framework: MemTropy}
Here, we summarize the structure and function of the simulation framework used for the calculations carried out in this manuscript. The simulation framework in its entirety with more detailed instructions on its use is available at \url{https://gitlab.com/taylor-schaffner/memtropic}. The framework consists of an engine written in C++ wrapped into Cython where it is configured and called. In it's current form, it may be imported to an interactive python session or executed as script. The full suite contains a user-friendly GUI with an on-going implementation of tools for exploring interaction networks among extended objects in a 2D membrane driven by entropy.

\section{Data and Analysis Files}
All post-simulation data analysis was carried out using MATLAB. Analysis code and data files can be found organized by figure at \url{https://gitlab.com/taylor-schaffner/ppi-paper-figure-data}.

\section{Detailed Balance on the Membrane}
Equilibration of the 2D Ising model membrane in our model is done in a manner that conserves the average spin value (magnetization) of the system. In other words, rather than flipping spins, spins on the lattice are exchanged. In each case, the detailed balance condition is satisfied by using the standard Metropolis acceptance probability
\beq
    \label{eq:metropolis_db}
    P(A \rightarrow B) = \text{Min}[1, \exp(-\beta(E_B - E_A))]
\eeq
Moves that do not increase energy are accepted, and moves that do are accepted according to a Boltzmann distribution. In each implementation case considered in the following, the same acceptance probability is used with only changes in the definitions of state $A$ and $B$.

\subsection{Local Kawasaki Algorithm}
Individual spins, here labeled lipids, migrate in our model by swapping with their neighbors according to \ref{eq:metropolis_db}. For a given move, the states $A$ and $B$ differ only by the $6$ modified bonds following the proposed exchange. In this way, lipids migrate in a way that conceptually replicates diffusion in the membrane.

\subsection{Non-local Kawasaki Algorithm}
Prior to the primary run-time in which protein-protein interaction network data is collected for a system, time to thermal equilibrium steady-state may be decreased considerably by first performing non-local spin swaps of the individual lipids. These moves similarly obey \ref{eq:metropolis_db} but in this case, the difference between states $A$ and $B$ consists of $8$ modified bonds.

\subsection{Diffusion of Membrane Inclusions}
Following an algorithm previously utilized for studies on lattice-based colloidal chemistry simulations \cite{Tasios2016}, we model diffusion of a membrane inclusion as a non-local, cluster spin exchange. After selecting a random inclusion (PPI component), one of four cardinal directions is randomly chosen, and an attempt is made to simultaneously swap the membrane spins adjacent to the inclusion's leading edge (outside the inclusion) with the corresponding spins on the back edge on the inclusion (inside inclusion). The difference between states $A$ and $B$ in this case is again the set of bonds altered--the number of which increases with disk radius.

\subsection{Thermal equilibration}
\subsubsection{Simulations with fixed order parameter}
In the majority of the results presented, the magnetization order parameter/fraction of dark spins is fixed for the runtime. In these systems, we ensure that the system is in thermal equilibrium before data collection. One means by which we achieve this is to equilibrate the membrane sans disks via non-local Kawasaki swaps, and then add the disks onto the lattice while minimizing order-parameter changes--i.e. by placing a dark phase-preferring disk into a pool of dark spins only.

\subsubsection{Simulations with disks changing partitioning}
In systems with an interaction network that modifies the target's partition preference, we can thermally equilibrate before hand, but the partition change will always break the equilibrium. Furthermore, while the observed phenomenon we present here and call 'pocketing' will occur from simulations initiated at thermal equilibrium, we note that its occurrence may be greatly increased by initializing the system with conducive thermal inhomogeneities. The results in the domain-changing section were initialized by first placing the disks on the lattice in a uniformly distributed manner and then performing non-local Kawasaki sweeps to attempt to equilibrate the surrounding solvent. The result is a phase-separated but segmented lattice that is highly conducive to forming pockets as many recruiters become isolated.

\section{Cross-Correlation Functions}
All correlation functions considered in the results presented are two-dimensional (cross) correlation functions computed as described below. The computational outline considers a general set of two signal matrices $A$ and $B$, and then the remaining sections define what those matrices contain in each case considered.

\subsection{Computation of 2D Cross-Correlations}
We begin with two signal matrices $A$ and $B$ and assume that they are square and share the same dimensions, $N \cross N$. We make use of discrete Fourier transform (DFT) techniques to efficiently compute two-dimensional cross-correlation functions.

\be
\I{zero-padding}

All data considered in this manuscript originate from systems with periodic boundary conditions. As a result, we may forgo zero-padding signal matrices before computing the correlations.

\I{zero-mean signal}

Now we define a set of shifted signal matrices such that each has a mean of zero.
$$A = A - \avg{A}$$
$$B = B - \avg{B}$$
With these definitions, we are ready to proceed with the correlation calculation.

\I{Fourier transform each signal}

Discrete Fourier transforms of $A=A(x,y)$ and $B=B(x,y)$ are computed to obtain the frequency space representations $A_k(p,q)$ and $B_k(p,q)$
$$A_k(p,q) = \sum_{x=0}^{N-1} \sum_{y=0}^{N-1} A(x,y) e^{-\frac{\pi i}{N} (px + qy) }$$
and similarly for $B_k(p,q)$.

\I{compute the power spectrum}

The power spectrum denoted $G_k(p,q)$ is then computed by element-wise multiplication between $A_k(p,q)$ and the complex conjugate of $B_k(p,q)$.
$$G_k(p,q) = A_k(p,q) B_k^*(p,q) $$

\I{un-normalized cross-correlations}

The power-spectrum is then inverse-Fourier transformed to obtain the un-normalized two dimensional cross-correlation function in position space $\tilde{G}(x,y)$

$$\tilde{G}(x,y) = \frac{1}{N^2} \sum_{p=0}^{N-1} \sum_{q=0}^{N-1} G_k(p,q) e^{\frac{\pi i}{N} (px + qy) }$$

\I{normalization}

To normalize the two dimensional cross-correlation function, we divide the result by the square root of the product of the squared-sum of each signal matrix, i.e. each density. 

$$ G(x,y) = \frac{\tilde{G}(x,y)}{\sqrt{ \sum_{x,y} A(x,y)^2 \sum_{x,y} B(x,y)^2}}$$

\ee

\subsection{Computation of Radial Average}
To more easily digest the information contained in the computed two-dimensional cross-correlation functions, we reduce the dimensionality by radially averaging.

\be
\I{shift zero-frequency to the center}

Before radially-averaging $G(x,y)$, we perform what is commonly referred to as a Fourier shift which involves rotating the data such that the interesting components are centered. For our two-dimensional, data this is simply a swap of the first quadrant with the third quadrant and a swap of the second quadrant with the fourth.

\I{radially average}

The shifted $G(x,y)$ is radially-averaged where the radius $r$ is defined as the distance from the center of the cross-correlation matrix. Note that our lattice (and hence signal matrices) is even dimensions and, in the discrete Fourier transform algorithms we use, $(0,0)$ is defined to be the spatial center. Thus following the Fourier shift, the new center is located at index $(N/2,N/2)$, setting $r=0$ for the radial average. This gives us $G(r)$.

\ee

\subsection{Protein-Protein Cross-Correlations}
In the case protein-protein cross-correlation calculations we consider in the main results the cross-correlation between the target component and its activator and between the target component and its inactivator. For instance, if we consider the activator-target cross-correlation specifically, matrix $A$ is then a matrix corresponding to the standard simulation lattice but with $1$'s at all sites containing an activating component and $0$'s elsewhere and matrix $B$ is a similar matrix with $1$'s and $0$'s denoting the location of all targets.

\subsection{Spin-Spin Correlations}
Spin-spin correlation functions are computed using the same procedure outlined above, but in such a calculation, $A = B$. Additionally, in this signal matrix, up spins are represented by $1$ while down spins are represented by $-1$.

\section{Varying Component Size}

\subsection{Fitting activity as a function of temperature for various PPI component sizes}
When comparing the activity response to changes in temperature for various PPI component sizes, we fit equation \ref{eq:sigmoid_eq} and obtained the following tabled fit values alongside the 95\% confident interval bounds for each value. All fits were performed in MATLAB.

\begin{center}
\begin{tabular}{||c c c c c||} 
 \hline
 Fit Parameter & $r = 0$ & $r=1$ & $r=2$ & $ r=3$ \\ [0.5ex] 
 \hline\hline
 $L$ & 0.292 (0.273,0.311) & 0.184 (0.168,0.200) & 0.202 (0.185,0.220) & 0.138 (0.124,0.152) \\ 
 \hline
 $R$ & 0.565 (0.551,0.579) & 0.737 (0.707,0.768) & 0.793 (0.770,0.816) & 0.760 (0.741,0.778) \\
 \hline
 $\tau_0^{\textrm{fit}}$ & 0.996 (0.993,0.999) & 1.014 (1.011,1.016) & 1.008 (1.007,1.010) & 1.005 (1.003,1.006) \\
 \hline
 $\Delta \tau$ & 0.0209 (0.0177,0.0257) & 0.0207 (0.0184,0.0236 )& 0.0165 (0.0147,0.0188) & 0.0150 (0.0137,0.0166) \\
 \hline
\end{tabular}
\end{center}

The coupling of inclusions (any disk with $r > 0$ in this context) to the 2D Ising membrane has the effect of making the system effectively \textit{colder} than without them. This perturbation in our system is viewed as an increase in the true critical temperature of the system $T_c$ to simplify the interpretation of $\tau$ throughout the manuscript. In the process of fitting the activity data in Fig. \ref{fig:disk_size}\textbf{A}, an estimate of this increased critical temperature $\tau_0^{\textrm{fit}}$ is calculated. This shift accounts for the $x$-axis shift in the activity curves. In practice, the approximate new critical point is fairly close to the original critical point provided that the PPI components occupy a reasonably small fraction (less than $ 2 \% $ for all cases considered here) of the total lattice. The exact numerical value of this shift is largely irrelevant to the qualitative arguments put forth in this work, so this discussion and the fit values are included primarily as a visual aid.

\subsection{Maximum slope of activity vs temperature data}

\figSUPPDISKSIZE

% \bibliography{references}

\twocolumn

\end{document}